\documentclass[prb,twocolumn,superscriptaddress,showpacs,preprintnumbers,amsmath,amssymb]{revtex4}
\usepackage{amssymb}

\usepackage{graphicx}% Include figure files
\usepackage{dcolumn}%Align table columns on decimal point
\usepackage{bm}% bold math
\usepackage{epsfig}
\newcommand{\ignore}[1]{}

\begin{document}
\title{Analytical Study of Electronic Structure in Armchair Graphene Nanoribbons}
\author{Huaixiu Zheng}
\address{Electrical and Computer Engineering, University of Alberta, AB T6G 2V4, Canada}%

\author{Zhengfei Wang}
\affiliation{Hefei National Laboratory for Physical Sciences at
Microscale, University of Science and Technology of China, Hefei,
Anhui 230026, People's Republic of China}

\author{Tao Luo}
\affiliation{Hefei National Laboratory for Physical Sciences at
Microscale, University of Science and Technology of China, Hefei,
Anhui 230026, People's Republic of China}

\author{Qinwei Shi}\thanks{Corresponding author. E-mail: phsqw@ustc.edu.cn}
\affiliation{Hefei National Laboratory for Physical Sciences at
Microscale, University of Science and Technology of China, Hefei,
Anhui 230026, People's Republic of China}

\author{Jie Chen}\thanks{Corresponding author. E-mail: jchen@ece.ualberta.ca}
\address{Electrical and Computer Engineering, University of Alberta, AB T6G 2V4, Canada}%

\begin{abstract}
We present the analytical solution of the wavefunction and energy
dispersion of armchair graphene nanoribbons (GNRs) based on the
tight-binding approximation. By imposing hard-wall boundary
condition, we find that the wavevector in the confined direction is
discretized. This discrete wavevector serves as the index of
different subbands. Our analytical solutions of wavefunction and
associated energy dispersion reproduce the numerical
tight-binding results and the solutions based on the $\bf k\cdot p$
approximation. In addition, we also find that all armchair GNRs with
edge deformation have energy gaps, which agrees with recently
reported first-principles calculations.


\end{abstract}

\maketitle

\section{Introduction}
Graphene, as a promising candidate of future nanoelectronic
components, has recently attracted intensive research
attention.{\cite{1,2,3,4,5,6}} Graphene consists of a single atomic
layer of graphite, which can also be viewed as a sheet of unrolled
carbon nanotube. Several anomalous phenomena ranging from
half-integer quantum Hall effect, non-zero Berry's phase {\cite{2}},
to minimum conductivity {\cite{3}}, have been observed in
experiments. These unusual transport properties may lead to novel
applications in carbon-based nanoelectronics. In addition, the
carriers in graphene behave as massless relativistic particles with an
effective `speed of light' $c_*  \approx 10^6 m/s$ within the
low-energy range ($\varepsilon  < 0.5eV$). {\cite{2}} These massless
Dirac fermions in graphene manifest various quantum electrodynamics
(QED) phenomena in the low energy range such as Klein paradox phenomenon.
{\cite{6}} Ribbons with a finite width of graphene, referred to as
graphene nanoribbons (GNRs), have also been studied extensively.
{\cite{JPSJ96,KNPRB96,KWPRB99,MEPRB06,LouiePRL06,LouieNat06,YMPRB99,LBPRB06}}
Recent experiments by using the mechanical method
{\cite{2,3}} and the epitaxial growth method {\cite{4,CBJPCB04}} show it is
possible to make GNRs with various widths.

\begin{figure}[htpb]
\begin{center}
\epsfig{figure=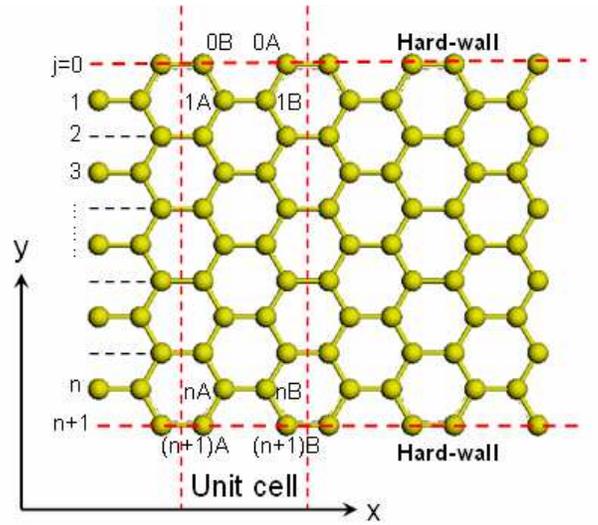,height=8cm,width=9cm}
\vspace{-0.4in}
\end{center}
\caption{(color online) Structure of an armchair graphene
nanoribbon, consisting of sublattices, A and B. The width of the
armchair GNR is $n$. Every unit cell contains $n$ number of A and B
sublattices. Two additional hard walls ($j=0, n+1$) are imposed on
both edges. } \label{fig:fig_1}
\vspace{-0.4in}
\end{figure}
The carbon atoms on the edge of GNRs have two typical topological
shapes, namely armchair and zigzag. The analytical wavefunction and
energy dispersion of zigzag nanoribbons have been derived by several
research groups. {\cite{KSJPSJ06,APL06}} For armchair GNRs, the
analytical forms of wavefunctions within the low energy range have
been worked out based on the effective-mass approximation.
{\cite{LBPRB06}} It is predicted that all zigzag GNRs are metallic
with localized states on the edges, {\cite{KNPRB96, KWPRB99,
KSJPSJ06, APL06}} while armchair GNRs are either metallic or
insulating, depending on their widths.
{\cite{JPSJ96,KNPRB96,KWPRB99,MEPRB06,KSJPSJ06,LBPRB06}} To-date, there is no
general expression of the wavefunction in armchair GNRs.
In this paper, we derive a general analytical expression
of wavefunction and eigen-energy in armchair GNRs applicable to various
energy ranges. In part II, we focus on perfect armchair GNRs
without any edge deformation and derive the energy dispersion by
imposing the hard-wall boundary condition. Due to the quantum
confinement, the spectrum breaks into a set of subbands and the
wavevector along the confined direction becomes discretized. We observe
that the electronic structure of perfect armchair GNRs strongly
depends on the width of the ribbon. The system, for instance, is metallic
when $n=3m+2$ and is insulating otherwise, where $m$ is an integer.
{\cite{JPSJ96, KNPRB96, KWPRB99, MEPRB06, KSJPSJ06,LBPRB06}}
Furthermore, we study the low energy electronic structure.
The linear dispersion relation is observed in armchair GNRs. In
part III, we evaluate the effect of deformations on the edges on the electronic structure of armchair GNRs.
Calculation results based on the derived analytical wavefunction
show that all armchair GNRs have nonzero energy gaps
due to the variation of hopping integral near the edges. This
observation is in line with the recently reported first-principle calculations.
{\cite{LouiePRL06}}

\section{Perfect armchair graphene nanoribbon}
The structure of armchair GNRs consists two types of
sublattices A and B as illustrated in Fig. 1. The unit cell
contains $n$ A-type atoms and $n$ B-type atoms. Based on the
translational invariance, we choose plane wave basis along the $x$
direction. Within the tight-binding model, the wavefunctions of A
and B sublattices can be written as
\begin{eqnarray}
\nonumber \left|\psi\right\rangle _A  = \frac{1}{{N_A
}}\sum\limits_{x_{A_i } } {\sum\limits_{i = 1}^n {e^{ik_x x_{A_i } }
} } \phi _A (i)\left| {A_i } \right\rangle  \\
\left|\psi\right\rangle _B  = \frac{1}{{N_B }}\sum\limits_{x_{B_i} }
{\sum\limits_{i = 1}^n {e^{ik_x x_{B_i } } } } \phi _B (i)\left|
{B_i } \right\rangle , \label{eqn:1}
\end{eqnarray}
where $\phi _A (i)$ and $\phi_B(i)$ are the components for A and B
sublattices in the $y$ direction, which is perpendicular to the
edge. $\left| {A_i } \right\rangle$ and $\left| {B_i }
\right\rangle$ are the wave functions of the $p_z$ orbit of a carbon
atom located at A and B sublattices, respectively. To solve $\phi _A
(i)$ and $\phi_B(i)$, we employ the hard-wall boundary condition
\begin{eqnarray}
\nonumber\phi _A (0) &=& \phi _B (0) = 0 \\
\phi _A (n + 1) &=& \phi _B (n + 1) =0.\label{eqn:2}
\end{eqnarray}
Choosing $\phi _A (i) = \phi _B (i) = \sin (\frac{{\sqrt{3}q_y
a}}{2}i)$ and substituting them into Eq. (2), we get
\begin{eqnarray}
q_y = \frac{2}{{\sqrt3a}}\frac{{p\pi }}{{n + 1}} , \ \ \ p = 1,2,
\cdots ,n.\label{eqn:3}
\end{eqnarray}
$q_y$ is the discretized wavevector in the $y$ direction and $a =
1.42\AA$ is the bond length between carbon atoms. To obtain the
normalized coefficients, $N_A$ and $N_B$, we introduce the
normalization condition
\begin{equation}
_A \left\langle \psi \right|\left. \psi \right\rangle _A = _B
\left\langle \psi \right|\left. \psi \right\rangle _B  = 1.\nonumber
\end{equation}
It is straightforward to obtain $N_A=N_B=\sqrt {\frac{N_x(n +
1)}{2}}$, where $N_x$ is the number of unit cells along the $x$
direction. The total wavefunction of the system can be constructed
by the linear combination of $\psi _A$ and $\psi_B$
\begin{eqnarray}
\nonumber\left| \psi  \right\rangle  = C_A (\sqrt {\frac{2}{{N_x (n
+ 1)}}} \sum\limits_{x_{A_i } } {\sum\limits_{i = 1}^n {e^{ik_x
x_{A_i } } }
} \sin (\frac{{\sqrt{3}q_y a}}{2}i)\left| {A_i } \right\rangle ) \\
+ C_B (\sqrt {\frac{2}{{N_x (n + 1)}}} \sum\limits_{x_{B_i } }
{\sum\limits_{i = 1}^n {e^{ik_x x_{B_i } } } } \sin (\frac{{\sqrt{3}q_y
a}}{2}i)\left| {B_i } \right\rangle ). \nonumber \\ \label{eqn:4}
\end{eqnarray}
Under the tight binding approximation, the Hamiltonian of the system is
\begin{eqnarray}
H=\sum\limits_i {\varepsilon _i } \left| i \right\rangle
\left\langle i \right|-t_{i,j}\sum\limits_{ < i,j > } {(\left| {i}
\right\rangle \left\langle {j} \right|)},\label{eqn:5}
\end{eqnarray}
where $\left\langle{i}, {j}\right\rangle$ denotes the nearest
neighbors.

In perfect armchair GNRs, we set $t_{i,j}=t$ and ${\varepsilon _i
}=\varepsilon$. By Substituting Eq. (4) and Eq. (5) into the
Schrodinger equation, we can easily obtain the following matrix
expression:
\begin{eqnarray}
\left( {\begin{array}{*{20}c}
   \varepsilon & \mu   \\
   {\mu ^* } & \varepsilon  \\
\end{array}} \right)\left( {\begin{array}{*{20}c}
   {C_A }  \\
   {C_B }  \\
\end{array}} \right) = E\left( {\begin{array}{*{20}c}
   {C_A }  \\
   {C_B }  \\
\end{array}} \right),\label{eqn:6}
\end{eqnarray}
where $ \mu  = _A \left\langle \psi  \right|H\left| \psi
\right\rangle _B = t(2e^{ik_x \frac{{a}}{2}}
\cos(\frac{{\sqrt3a}}{2}q_y)  + e^{ - ik_x a} )$. Solving Eq. (6),
we get the energy dispersion and wavefunction as
\begin{equation}
\nonumber E =\varepsilon  \pm \left|\mu \right|,
\end{equation}
\begin{eqnarray}
\left|\psi\right\rangle_\pm=\frac{{\sqrt2}}{2}(\left|\psi
\right\rangle_A\pm \sqrt {\frac{{\mu ^* }}{\mu }}\left|\psi
\right\rangle_B).\label{eqn:8}
\end{eqnarray}
Here, $\pm$ denotes the conduction and valance bands, respectively.
$- \frac{\pi }{2} \le \frac{{3k_x a}}{2} \le \frac{\pi }{2}$ is
required within the first Brillion zone. These results are valid for
various energy ranges.
\begin{figure}
\begin{center}
\epsfig{figure=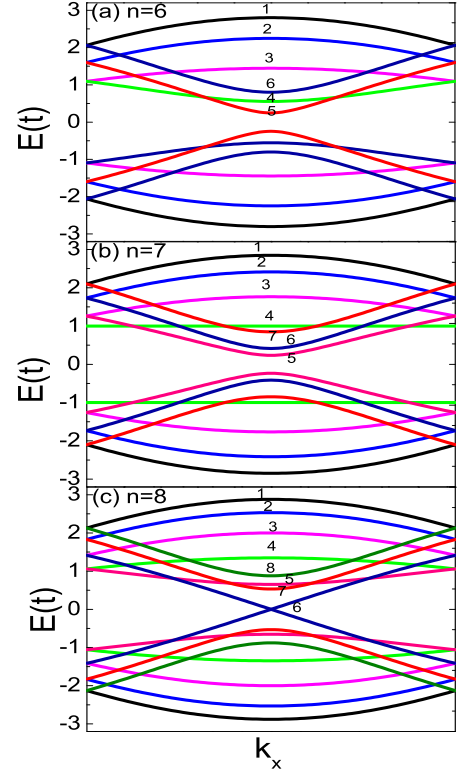,height=12cm,width=8.0cm}
\vspace{-0.5in}
\end{center}
\caption{(color online) Electronic structures of perfect armchair
GNRs with various widths, (a) n=6 (b) 7 and (c) 8, respectively.  The wavevector is
normalized based on the primitive translation vector of individual
GNRs. The value of $p$ for each subband is labeled in the figure.}
\label{fig:fig_2}
\vspace{-0.2in}
\end{figure}

\begin{figure}
\begin{center}
\epsfig{figure=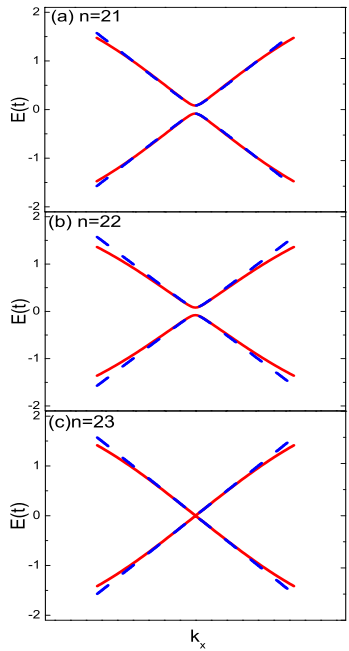,height=10cm,width=6.2 cm}
\vspace{-0.4in}
\end{center}
\caption{(color online) The first conductance and valence bands
within the first Brillion zone: exact solutions from Eq. (7) (red
solid line) and low-energy approximation from Eq. (10) (blue dash
line) for armchair GNRs with various widths,  (a) $n=21$ (b) $n=22$ (c)
$n=23$, respectively. The wavevector is normalized based on the primitive
translation vector of individual GNRs.}
\vspace{-0.2in}
\end{figure}

Fig.~\ref{fig:fig_2} shows the energy dispersion for perfect armchair GNRs with
width $n=6, 7$ and $8$. Here, we set $\varepsilon=0$. The results are the same as those obtained by using the
numerical tight-binding method. The electronic structures of
armchair GNRs depend strongly on their widths. When $n=8$, the
lowest conduction band and the upmost valence band touch at the
Dirac point, which leads to the metallic behavior of $n=8$ armchair
GNRs. Armchair GNRs, however, are insulating when $n=6$ and $n=7$.
Armchair GNRs with the width of $n=3m+2$ ($m$ is an integer)
 are generally metallic and otherwise are insulating. {\cite{KNPRB96,LBPRB06}}
In addition, we observe several interesting features in the band
structures of armchair GNRs.\\

(i) If $n$=7, a flat conduction/valence band ($p=4$) exists
as shown in Fig.~\ref{fig:fig_2} (b). Such a flat
band generally corresponds to $\frac{{p}}{{n + 1}} = \frac{1}{2}$ or
equivalently $\cos\frac{{p\pi }}{{n + 1}}=0$. The energy dispersion
becomes independent of $k_x$ and the eigen-energy always equals $\pm
\left| t \right|$. Flat band, in general,  exists only when $n$ is odd. \\

(ii) The subbands can be labeled by the quantum number $p$. Combined
with the wave number $k_x$ along the $x$ direction, the quantum
number $p$ can be used to define the chirality of the electrons in
quasi-1D graphene ribbons similar to that in 2D graphene. To
identify different subbands, we need the the quantum number $p_i$ of
the $i$th conduction/valence band. Here, the definition of the
sequence of subbands is referred to as the value of eigen-energy, $E_C$, in
the center of first Brillion zone ($k_x=0$),
\begin{eqnarray}
E_C=\pm t\left| {2\cos \frac{{p\pi }}{{n + 1}} +1} \right|.
\end{eqnarray}

For the metallic armchair GNRs with width $n=3m+2$ when
$\frac{{p\pi }}{{n + 1}} = \frac{2\pi}{3}$ or equivalently $p=2m+2$,
the energy gap between conduction and valence bands is zero.
Therefore, $p_1=2m+2$ corresponds to the first conduction/valence
band in $n=3m+2$ GNRs. For the second conduction/valence bands,
$E_C$ should have the minimal nonzero value compared to the third or
even higher band. After analyzing the value of $E_C$, we find that
$p_2=2m+3$, $p_3=2m+1$ for metallic armchair GNRs ($n=3m+2$). By
similar analysis, for $n>10$, we can obtain $p_1=2m+1$, $p_2=2m$,
$p_3=2m+2$ for $n=3m$ armchair GNRs and $p_1=2m+1$, $p_2=2m+2$,
$p_3=2m$  for $n=3m+1$ armchair GNRs, respectively. For all
subbands, there is no general rule of the subband index $p$.\\

(iii) Lots of research interest have been focusing on
the energy dispersion and wavefunction of 2D graphene and 1D GNRs within
the low-energy range. {\cite{2,
LouiePRL06, LBPRB06}} Low-energy electrons behaves as massless relativistic particles in a
2D infinite graphene system. {\cite{1, 2, 3, 6, LBPRB06}} Whether electrons keep their
relativistic property when they are confined in quasi-1D graphene
nanoribbons is an interesting issue. In what follows, we will focus
on the expansion of our analytical expressions to the low energy
limit. When $\frac{{p\pi }}{{n + 1}} \to \frac{2}{3}\pi$ and
$\frac{{3k_x a}}{2}
\to 0$, we rewrite the eigenenergy in Eq. (7) as\\
\begin{eqnarray}
E \approx  \pm {\textstyle{3 \over 2}}at\sqrt {k_x^2  + \tilde q_y^2
} \approx \pm \hbar v_F k ,
\end{eqnarray}
where $\tilde q_y (p) = \frac{2}{{\sqrt 3 a}}(\frac{{p\pi }}{{n +
1}} - \frac{2}{3}\pi )$, $p$ is the subband index. This low energy
expansion generates the $E\propto k$ linear dispersion, with Fermi
velocity $v_F \approx 10^6 m/s$. This expression reproduces the
result of $\bf k \cdot p$ approximation. {\cite{LBPRB06}} Note that
the wavevector in the confined direction ($\tilde q_y$) is
discretized, corresponding to different subbands. What is worthy
mentioning is that this energy dispersion works well only at the
low-energy limit. By substituting the value of $p_1$ into Eq. (9), we
get the low energy expansion of the first conduction/valence band
for armchair GNRs as
\begin{eqnarray}
\nonumber E_1 (3m)\mathop  \approx \limits^{k_x  \to 0}  &\pm& {\textstyle{3 \over 2}}at\sqrt {k_x^2  + (\frac{{2\pi }}{{3\sqrt 3 (3m + 1)a}})^2 }  \\
\nonumber E_1 (3m + 1)\mathop  \approx \limits^{k_x  \to 0}  &\pm& {\textstyle{3 \over 2}}at\sqrt {k_x^2  + (\frac{{2\pi }}{{3\sqrt 3 (3m + 2)a}})^2 }  \\
E_1 (3m + 2)\mathop  \approx \limits^{k_x  \to 0} &\pm&
\frac{{3at}}{2}k_x.
\end{eqnarray}

\begin{figure}
\begin{center}
\epsfig{figure=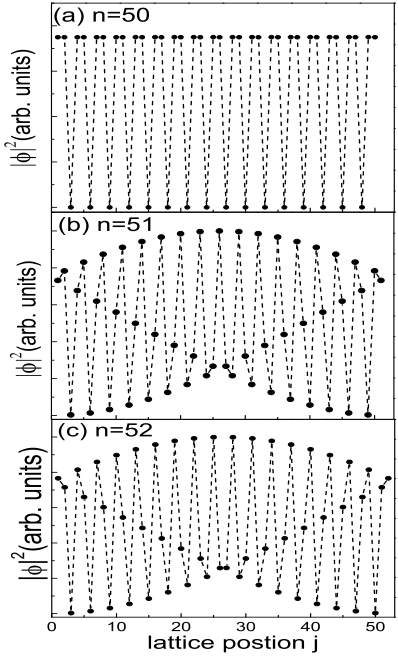,height=10cm,width=7.0cm}
\vspace{-0.4in}
\end{center}
\caption{Local density of the states in the first conduction/valence
band at $k_x=0$ for armchair GNRs with various widths, (a) $n$=50 (b) $n=51$
and (c) $n=52$, respectively.
(These values of $n$ are so chosen to match the results in reference {\cite{LBPRB06})}} \label{fig:fig_4}
\vspace{-0.2in}
\end{figure}
Fig. 3 shows the quality of low energy approximation. For large
width armchair GNRs, low energy approximation seems work well except
the edge of first Brillion Zone. As the width gets larger, the
quantum confinement due to the edge becomes less important and the
1D nanoribbons tend to behave like 2D graphene. For large $n$, as expected, the
band structure generates the linear dispersion relationship, $E\propto k_x$, in the low-energy limit.

In addition, from the expression of wavefunction, we also obtain the
local density of electronic states in perfect armchair GNRs, $ P_A
(i) = P_B (i) \propto \sin ^2 (\frac{{p\pi }}{{n + 1}}i)$ . Fig. 4
shows the squared wave functions of the lowest conduction band at
the center of first Brillion Zone. Note that Fig. 4(a) and (c)
reproduce the results of the $\bf k\cdot p$ approximation.
{\cite{LBPRB06}} The state density oscillates as a function of the
lattice position. The oscillation period is related to $\frac{{n +
1}}{p}$. For $n=3m+2$ armchair GNRs, the oscillation period is just
3, which is shown clearly in Fig. 4(a). For $n=3m$, $3m+1$ armchair
GNRs, we should write $\frac{{n + 1}}{p}$ into an irreducible form
$\frac{\alpha}{\beta}$. The oscillation period then equals $\alpha$, which is
the numerator of the irreducible form of $\frac{{n + 1}}{p}$.
To match the results presented in {\cite{LBPRB06}},
we choose $n=51$ and $n=52$ as an example. We get $i=51$ and $i=52$, respectively. As shown in Fig. 4(b)
and (c), the oscillation period of state density for $n=51$ and
$n=52$ armchair GNRs equals their width.

\section{Energy Gap and Wavefunction for Edge-deformed GNRs}
Because every atom on the edge has one dangling bond unsaturated,
the characteristics of the $C-C$ bonds at the edges can change GNRs' electronic structure
dramatically. {\cite{MEPRB06, TKPRB00}} To determine the band gaps
of GNRs on the scale of nanometer, edge effects should be considered
carefully. The change of edge bonds length and angle can lead
to considerable variations of electronic structure, especially
within the low energy range. {\cite{LouiePRL06, LouieNat06}} In
previously reported work, the edge carbon atoms of GNRs are all passivated by
hydrogen atoms or other kinds of atoms/molecules. {\cite{LBPRB06,
MEPRB06, TKPRB00, LouiePRL06, LouieNat06}} The bonds between
hydrogen and carbon are different from those $C-C$ bonds. Accordingly,
the transfer integer of the $C-H$ bonds and on-site energy of carbon
atoms at the edges are expected to differ from those in the
middle of GNRs. The bond lengths between carbon atoms at the edges
are predicted to vary about $3\%-4\%$ when hydrogenerated.
{\cite{LouiePRL06}} Correspondingly, the hopping integral increases
about $12\%$ extracted from the analytical tight binding expression.
{\cite{DPPRB95, LouiePRL06}} To evaluate the effect of various kinds
of edge deformation, we carried out general theoretical calculation
and analysis with our analytical solution of armchair GNRs.
In general, we can set the variation of the transfer integer and on-site
energy as $\delta t_{i,j}$, $\varepsilon _i$ for the $i$th A-type
or B-type carbon atom in the unit cell. The Hamiltonian of the
GNRs with deformation on the edge can be rewritten as
\begin{eqnarray}
H = \varepsilon _i \sum\limits_i {\left| i \right\rangle }
\left\langle i \right| - \sum\limits_{ < i,j > }^{} {(t + \delta
t_{i,j} )} \left| i \right\rangle \left\langle j \right|
\end{eqnarray}
It is readily to obtain the energy dispersion and wavefunction by
solving the Schrodinger equation
\begin{eqnarray}
\nonumber E = \gamma  \pm \left| {\mu  + \delta \mu } \right|
\end{eqnarray}
\begin{eqnarray}
\left| \psi  \right\rangle _ \pm   = \frac{{\sqrt 2 }}{2}(\left|
\psi  \right\rangle _A  \pm \sqrt {\frac{{(\mu  + \delta \mu )^*
}}{{\mu  + \delta \mu }}} \left| \psi  \right\rangle _B ),
\end{eqnarray}
where $\gamma  = \frac{2}{{n + 1}}\sum\limits_{i = 1}^n {\varepsilon
_i } \sin ^2 (\frac{{p\pi }}{{n + 1}}i)$ is the energy shift
originating from the variation of on-site energy, while the energy
shift from the hopping integral variation is {\large
\begin{eqnarray}
\delta \mu  =  - \frac{{2t}}{{n + 1}}\sum\limits_{i = 1}^n [\delta
t_{i(A)i(B)} \sin ^2 (\frac{{p\pi }}{{n + 1}}i)e^{ - ik_x
a} \nonumber \\
 + \delta t_{i(A)i - 1(B)} \sin (\frac{{p\pi }}{{n + 1}}i)\sin
(\frac{{p\pi }}{{n + 1}}(i - 1))e^{\frac{{ik_x a}}{2}} \nonumber \\
 + \delta
t_{i(A)i + 1(B)} \sin (\frac{{p\pi }}{{n + 1}}i)\sin (\frac{{p\pi
}}{{n + 1}}(i + 1))e^{\frac{{ik_x a}}{2}} ]. \nonumber \\
\end{eqnarray}
}

Such a general expression could include various kinds of edge
deformations, ranging from the quantum confinement effect due to the
finite width, to the effect of saturated atoms or molecules attached
to edge carbon atoms. This result shows that the
deformation leads to a considerable deviation of the energy
dispersion relation and wavefunction of the deformed system from
those in perfect armchair GNRs. The local state density on both
kinds of sublattices, however, remains the same as that in perfect
armchair GNRs. The reason is that the wavefunctions of sublattices A
and B change their relative phases, but keep the magnitudes
unchanged. The variations from both the on-site energy and hopping
integral contribute to the energy shift, while, the change of
on-site energy has no contribution to the wavefunction as shown in
Eq. (8). To verify our findings, we reproduce the energy gap
observed in the recent work {\cite{LouiePRL06}} by considering
only the variation of hopping integrals of the bonds on the edges ($
\delta t_{11} = \delta t_{nn} = \delta t$, others equals zero).
The corresponding energy gaps for different width ribbons are as
follows:
\begin{eqnarray}
\nonumber \Delta_{3m}&=&\Delta_{3m}^0 -\frac{{8\delta t}}{{3m + 1}}\sin ^2 \frac{{m\pi }}{{3m + 1}}, \\
\nonumber \Delta _{3m + 1}&=&\Delta _{3m + 1} ^0  + \frac{{8\delta t}}{{3m + 2}}\sin ^2 \frac{{(m + 1)\pi }}{{3m + 2}}, \\
\Delta _{3m + 2}&=&\Delta _{3m + 2} ^0  + \frac{{2\delta t}}{{m +
1}},
\end{eqnarray}
where $\Delta_{3m}$, $\Delta _{3m + 1}$ and $\Delta _{3m + 2}$ are
the energy gaps of perfect armchair GNRs. Their values can be
extracted from Eq.(8): $2t\left| {2\cos \frac{{(2m+1)\pi }}{{3m+ 1}}
+1} \right|$, $2t\left| {2\cos \frac{{(2m+1)\pi }}{{3m+2}} +1}
\right|$ and $0$. This result suggests all armchair graphene ribbons
with edge deformation have nonzero energy gaps and are insulating
correspondingly.

\section{Conclusion}
In this paper, we study the electronic states of armchair
GNRs analytically. By imposing the hard-wall boundary condition, we find
the analytical solution of wavefunction and energy dispersion in
armchair GNRs based on the tight-binding approximation. Our results
reproduce the numerical tight-binding calculation results and the solutions using
the effective-mass approximation. We also derive the low-energy
approximation of the energy dispersion, which matches the
exact solution except for the edge of first Brillion zone. The linear
energy dispersion is observed  in armchair GNRs in the low energy
limit. In addition, we also evaluate the impact of the edge
deformation on GNRs and derive a general expression of wavefunction and
energy dispersion.
We can reproduce the energy gap for hydrogenerated armchair GNRs presented in {\cite{LouiePRL06}}.
When we consider the edge deformation, all armchair GNRs have nonzero energy gaps and
thus are insulting. Overall, the derived analytical form of the wavefunction can be used to
quantitatively investigate and predict various properties in
armchair graphene ribbons.
\vspace{-0.3in}

\section*{ACKNOWLEDGMENT}
This work is partially supported by the National Natural Science
Foundation of China under Grant no.10274076 and by National Key
Basic Research Program under Grant No.2006CB0L1200. Jie Chen would
like to acknowledge the funding support from the Discovery program
of Natural Sciences and Engineering Research Council of Canada (No.
245680). We also would like to thank Nathanael Wu and Stephen
Thornhill for their assistance with the finalization of this paper.

\end{document}